\documentclass[runningheads]{llncs}
\usepackage[T1]{fontenc}
\usepackage{newtxtext}       %
\usepackage[varvw]{newtxmath}       

\usepackage{graphicx}
\usepackage{hyperref}
\usepackage{multirow}
\usepackage{booktabs}
\usepackage{subcaption}
\usepackage{tcolorbox}
\usepackage{color}

\newcommand{\orcid}[1]{\href{https://orcid.org/#1}{\includegraphics[width=8pt]{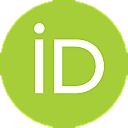}}}
\begin{document}
\title{A Comparative Reproducibility Study of Jev and Leading Reranking Models}
%
%
\author{Sajad Ebrahimi$^*$ \inst{1} \orcid{0009-0003-1630-3938} \and
Alireza Mirrokni$^*$ \inst{1} \orcid{0009-0002-9550-7017} \and
Negar Arabzadeh\inst{2}\orcid{0000-0002-4411-7089} \and
Ebrahim Bagheri\inst{1}\orcid{0000-0002-5148-6237}}
\authorrunning{S. Ebrahimi, A. Mirrokni et al.}

\institute{University of Toronto \and
University of California, Berkeley \\
\email{s.ebrahimi@utoronto.ca}}
\maketitle              
\def\thefootnote{*}\footnotetext{These authors contributed equally to this work.}

\begin{abstract}
Large Language Models achieve state-of-the-art effectiveness in zero-shot document reranking, but their reliance on autoregressive text generation makes them prohibitively slow and expensive for production search environments. Non-generative decision models, which return calibrated probabilities over fixed answer spaces without generating text, offer a promising alternative. In this large-scale reproducibility study, we benchmark Jev, a novel decision model, against established neural rerankers and generative LLMs across 14 diverse datasets spanning web passage search (TREC DL), domain-specific retrieval (BEIR), and reasoning-intensive retrieval (BRIGHT). We reproduce established relevance evaluation paradigms (pointwise binary, pointwise graded, and list-based selection) and map them to Jev's typed primitives. We find that Jev's pointwise graded mode (Score) achieves effectiveness comparable to GPT-4o-mini while delivering up to a 28$\times$ speedup and an 80\% cost reduction. Conversely, we demonstrate that list-based selection (Choice) fails catastrophically on datasets with long-tail passage distributions for both Jev and generative models, exposing fundamental vulnerabilities in listwise paradigms over long contexts. We release all code, prompts, and run files to support future reproducibility at \url{https://github.com/alirezamirrokni/JevRerank}.

\keywords{Document Reranking \and Generative Reranking \and System One Models}
\end{abstract}

\section{Introduction}

The integration of Large Language Models (LLMs) into information retrieval pipelines has established a new state-of-the-art for zero-shot document reranking. Researchers have developed a variety of prompting paradigms to extract ranking signals from LLMs, including pointwise relevance generation \cite{liang2023holistic,ma2023fine,bge_embedding}, pairwise comparisons \cite{qin2024large}, listwise sorting \cite{sun2023chatgpt}, and setwise selection \cite{zhuang2024setwise}. However, these approaches share a fundamental bottleneck in autoregressive text generation \cite{chen2024accelerating}. Decoding tokens sequentially is inherently slow, parsing generated text into structured ranking scores is brittle, and API costs scale linearly with both the massive input context and the number of generated tokens \cite{abdallah2026llm}. Consequently, deploying frontier LLMs as rerankers remains prohibitively expensive and slow for latency-sensitive search applications \cite{zhuang2024setwise,podolak2025beyond}.

Recently, a new class of non-generative decision models has emerged as a potential solution. Models like Jev (developed by TypeSafe AI) are trained to return calibrated probabilities over a fixed set of typed primitives, such as binary yes/no decisions, graded scores, or multiple-choice selections, without generating any text. By eliminating autoregressive decoding, these models promise the reasoning capabilities of large models at a fraction of the latency and cost. As such, we would like to explore \emph{whether a non-generative decision model can be a viable and efficient backbone for zero-shot document reranking.}

To answer this, we present a large-scale reproducibility study and benchmark of document rerankers across 14 diverse datasets, encompassing web passage search (TREC DL), heterogeneous domain-specific retrieval (BEIR), and reasoning-intensive retrieval (BRIGHT). \emph{While we reproduce the results of established state-of-the-art neural and generative LLM baselines on datasets not covered in their original papers}, our primary focus is evaluating the viability of non-generative models by benchmarking Jev. We map three established generative relevance evaluation paradigms to Jev's typed primitives (\texttt{Noul}, \texttt{Score}, and \texttt{Choice}) and compare its effectiveness and efficiency against neural models and frontier generative LLMs receiving the exact same instructions.

Our findings reveal clear differences between these paradigms. By mapping Jev's distinct typed modes to established zero-shot generative paradigms, we establish a fair playground to compare models with heterogeneous output types. Jev's pointwise graded mode (\texttt{Score}) proves highly effective, matching or exceeding the much more expensive GPT-4o-mini baseline on many datasets while offering massive latency and cost reductions. In contrast, list-based selection (\texttt{Choice}) struggles severely with both effectiveness and context limits across both Jev and generative models, highlighting the vulnerability of listwise paradigms to the lost-in-the-middle phenomenon and long-tail passage distributions.

We structure our investigation around the following research questions:
\begin{itemize}
    \item[] \textbf{RQ1.} How do non-generative decision models, supervised neural baselines, and frontier generative LLMs compare in terms of document reranking effectiveness?
    \item[] \textbf{RQ2.} How do different relevance evaluation paradigms (pointwise binary, pointwise graded, listwise selection) affect the effectiveness of a non-generative decision model?
    \item[] \textbf{RQ3.} How efficient is Jev compared to neural rerankers and generative LLMs?
\end{itemize}

In addressing these questions, we make the following \textbf{key contributions}: \textbf{(1)} An extensive reproduction of state-of-the-art neural and generative LLM baselines across 14 diverse datasets; \textbf{(2)} A mapping of generative relevance evaluation paradigms to non-generative decision models, enabling a systematic comparison against neural and generative baselines; \textbf{(3)} An efficiency analysis demonstrating the significant latency and cost advantages of non-generative models over autoregressive LLMs; \textbf{(4)} A failure case analysis of list-based selection paradigms, exposing their effectiveness degradation and context-limit vulnerabilities on long documents; and \textbf{(5)} The public release of all code, prompts, and run files to ensure full reproducibility, available at \url{https://github.com/alirezamirrokni/JevRerank}.
\section{Background and Related Work}

\subsection{Document Relevance Evaluation Paradigms}
The comparison paradigms used to evaluate document relevance originated in the Learning to Rank (LTR) literature \cite{liu2009learning}. With the advent of LLMs, these classical paradigms have been adapted into distinct generative prompting paradigms. Each adaptation trades off effectiveness against computational cost and context window constraints.

\textbf{Pointwise Reranking.} In classical LTR, pointwise methods score each document independently based on its absolute relevance to the query \cite{liu2009learning}. This paradigm has been adapted for LLMs by prompting the model to generate a relevance score for a single query-document pair. Early zero-shot methods prompt an LLM to output ``yes'' or ``no'', using the log-probability of the ``yes'' token as the ranking score \cite{liang2023holistic}. Recent advancements have shifted toward fine-grained graded relevance, where LLMs are prompted to assign a grade (e.g., 0 to 3) based on detailed rubrics, significantly improving effectiveness over binary judgments \cite{upadhyay2024umbrela}.

\textbf{Pairwise Reranking.} Pairwise LTR methods learn to predict the relative preference between two documents \cite{liu2009learning}. In the LLM context, pairwise prompting \cite{qin2024large} asks the model to compare two documents side-by-side and select the more relevant one. While highly effective, this approach requires $O(k^2)$ inferences to sort $k$ documents, making it computationally prohibitive for large candidate lists without complex pruning heuristics. 

\textbf{Listwise and Setwise Reranking.} Listwise LTR directly optimizes the ordering of an entire list of candidate documents \cite{cao2007learning}. To adapt this for LLMs and reduce inference calls, listwise prompting approaches like RankGPT \cite{sun2023chatgpt} and RankZephyr \cite{pradeep2023rankzephyr} insert multiple documents into a single prompt and ask the LLM to output a sorted list of document identifiers. Because context limits and the lost-in-the-middle phenomenon \cite{liu2024lost,bai2024longbench} degrade performance on large lists, these methods typically employ sliding window strategies. Alternatively, Setwise prompting \cite{zhuang2024setwise} asks the LLM to select only the \textit{single most relevant} document from a set, using sorting algorithms like Heapsort to construct the final ranking. Recent reproducibility studies have extensively analyzed Setwise approaches, highlighting their efficiency gains \cite{podolak2025beyond} and the impact of document truncation on their effectiveness \cite{lawrie2025reproducibility,meng2024ranked,meng2026re}. 

\subsection{Supervised Neural Rerankers}
While zero-shot LLMs dominate effectiveness benchmarks, supervised neural models remain the standard for production deployments due to their efficiency. Cross-encoders, such as those based on MiniLM \cite{wang2020minilm} or monoBERT \cite{nogueira2019passage}, concatenate the query and document and process them through a Transformer encoder. Sequence-to-sequence models like monoT5 \cite{nogueira2020document} are fine-tuned on MS MARCO to generate relevance tokens. Other efficient architectures include late-interaction models like ColBERT \cite{khattab2020colbert} and sparse expansions like SPLADE \cite{formal2021splade}. More recently, models like Qwen3-Reranker \cite{zhang2025qwen3} and RankZephyr \cite{pradeep2023rankzephyr} have distilled the reasoning capabilities of larger LLMs into smaller, more efficient architectures. While prior literature frequently features partial comparisons between scattered subsets of these neural models and generative LLMs, evaluations are often fragmented across different papers, datasets, and relevance evaluation paradigms. In this work, we provide a comprehensive, unified reproduction of these established supervised models and generative baselines across 14 diverse datasets, and we introduce Jev on top of this rigorous foundation to contextualize its performance.

\subsection{Efficiency in LLM Reranking}
As LLM rerankers scale, their computational overhead has become a primary research focus. Abdallah et al. \cite{abdallah2026llm} conducted a comprehensive reproducibility study on the efficiency and reasoning overhead of LLM-based retrievers, particularly on reasoning-intensive datasets like BRIGHT \cite{su2025bright}. Other efforts to accelerate LLM ranking include single-token decoding strategies like FIRST \cite{chen2024accelerating}, layer-wise attention analysis \cite{chen2026selectiveicr}, and integrating rerankers into deep-research agents \cite{meng2026revisiting}. Our work extends this line of inquiry by evaluating whether non-generative decision models can bypass the autoregressive bottleneck entirely.

\section{Jev and Its Typed Primitives}

\subsection{The Jev Decision Model}
Jev is a proprietary, API-only ``System One'' decision model. Unlike standard generative LLMs that decode text sequentially, Jev is trained via Reinforcement Learning for Calibrated Decisions (RLCD) to process an input state and directly return structured JSON containing calibrated probabilities over predefined typed primitives.

For our experiments, we use \texttt{jev-1.13.0} and verify the returned model ID on every API call to prevent version drift. The model accepts text, JSON objects, and arrays. It imposes a default context limit of 64k tokens per request and 32k tokens per individual question. Because TypeSafe AI does not provide a public tokenizer, request sizes must be estimated using byte heuristics. Jev charges \$0.042 per 1M input tokens, with output tokens provided for free.

\subsection{Mapping Generative Prompts to Typed Primitives}
To evaluate Jev, we map three established generative relevance evaluation paradigms onto its three core typed primitives: \texttt{Noul}, \texttt{Score}, and \texttt{Choice}, as illustrated in Figure~\ref{fig:prompts}. For every request, we construct a JSON state containing the query and an array of candidate passages. Questions reference these passages by their array indices. We rank the candidates based on the probabilities returned by the API, breaking any ties using the original BM25 rank.

\textbf{Pointwise Binary Relevance.} The \texttt{Noul} primitive returns a single probability in $[0, 1]$ indicating whether the answer to a question is ``yes''. We map this to the pointwise binary relevance paradigm \cite{liang2023holistic}. We pack 20 passages into a single request and ask 20 independent \texttt{Noul} questions formatted as ``Does passage[i] answer the query?''. We rank the passages in descending order of $P(\text{yes})$.

\textbf{Pointwise Graded Relevance.} The \texttt{Score} primitive evaluates an item against an ordered set of levels and returns a probability-weighted expected value. We map this to the pointwise graded relevance paradigm. We define four relevance levels (0 to 3) using the standard grading definitions from the TREC Deep Learning track \cite{craswell2020overview}, which were popularized for LLM reranking by UMBRELA \cite{upadhyay2024umbrela}. Similar to \texttt{Noul}, we pack 20 passages per request and ask 20 independent \texttt{Score} questions. We rank the passages by the expected grade returned by Jev.

\textbf{List-based Selection.} The \texttt{Choice} primitive selects one option from a provided list, returning a probability distribution over all options that sums to 1. We map this to the listwise/setwise prompting paradigm \cite{zhuang2024setwise}. We send a single request containing all 100 candidate passages (labeled \texttt{p0} to \texttt{p99}) and ask Jev to identify the most relevant passage. We rank the entire list based on the probability mass assigned to each passage.

\begin{figure}[t]
\centering
\begin{subfigure}[b]{0.95\textwidth}
\begin{tcolorbox}[colback=gray!5!white, colframe=gray!60!black, arc=2mm, boxrule=0.5pt, left=6pt, right=6pt, top=4pt, bottom=4pt]
\small
\textbf{Instruction:} Does \texttt{passages[i]} answer \texttt{query}? \\
\textbf{Criteria:} \\
\textbf{true:} The passage contains an answer to the query \\
\textbf{false:} The passage does not answer the query, even if it seems related
\end{tcolorbox}
\vspace{-1.5ex}
\caption{\texttt{Noul}}
\end{subfigure}

\vspace{2ex}
\begin{subfigure}[b]{0.95\textwidth}
\begin{tcolorbox}[colback=gray!5!white, colframe=gray!60!black, arc=2mm, boxrule=0.5pt, left=6pt, right=6pt, top=4pt, bottom=4pt]
\small
\textbf{Instruction:} How relevant is \texttt{passages[i]} to \texttt{query}? Consider the underlying intent of the search. \\
\textbf{Criteria:} \\
\textbf{Level 0:} The passage has nothing to do with the query \\
\textbf{Level 1:} The passage seems related to the query but does not answer it \\
\textbf{Level 2:} The passage has some answer for the query, but the answer may be unclear or hidden amongst extraneous information \\
\textbf{Level 3:} The passage is dedicated to the query and contains the exact answer
\end{tcolorbox}
\vspace{-1.5ex}
\caption{\texttt{Score}}
\end{subfigure}

\vspace{2ex}
\begin{subfigure}[b]{0.95\textwidth}
\begin{tcolorbox}[colback=gray!5!white, colframe=gray!60!black, arc=2mm, boxrule=0.5pt, left=6pt, right=6pt, top=4pt, bottom=4pt]
\small
\textbf{Instruction:} Select the passage in \texttt{passages} that is the most relevant one to \texttt{query}. \\
\textbf{Criteria:} \\
\textbf{p[i]:} The passage at \texttt{passages[i]}
\end{tcolorbox}
\vspace{-1.5ex}
\caption{\texttt{Choice}}
\end{subfigure}
\caption{Instructions and criteria sent to Jev across its three typed primitives: (a) \texttt{Noul} for pointwise binary relevance, (b) \texttt{Score} for pointwise graded relevance (using TREC DL definitions \cite{craswell2020overview,upadhyay2024umbrela}), and (c) \texttt{Choice} for list-based selection. The full prompts match the reproduced generative baselines exactly.}
\label{fig:prompts}
\vspace{-1em}
\end{figure}

\subsection{Handling Context Limits}
A fundamental challenge in zero-shot reranking is fitting candidate passages within the model's context window. Because Jev lacks a public tokenizer, we estimate request sizes using a conservative byte-budget heuristic (90,000 UTF-8 bytes). For pointwise modes, context limits are easily managed by packing 20 passages per request and recursively splitting the batch into smaller chunks if an API rejection occurs.

However, list-based selection requires evaluating candidates simultaneously. Passing the top 100 candidates in a single request frequently exceeds Jev's 64k token limit, particularly on datasets with long documents. To investigate how best to overcome this limitation, we adapted three engineering strategies from the LLM literature \cite{sun2023chatgpt,zhuang2024setwise} to serve as fallbacks when a native request fails:
\begin{enumerate}
    \item \textbf{Truncate:} A naive baseline where we iteratively apply a uniform character cap to all passages, shrinking the cap by a factor of 0.75 until the 100-passage request fits.
    \item \textbf{Sliding Window:} Adapted from RankGPT \cite{sun2023chatgpt}, we process the list in overlapping chunks (window size $w=50$, stride $s=25$), bubbling the most probable passages from bottom to top.
    \item \textbf{Tournament:} Inspired by Setwise prompting \cite{zhuang2024setwise}, we divide the 100 passages into groups of size $w$, select the top $m$ passages from each group, and run a final \texttt{Choice} request over the aggregated winners.
\end{enumerate}
In our main effectiveness tables, we report the performance of the ``Native'' \texttt{Choice} mode, which simply defaults to the BM25 ordering if the context limit is exceeded. We evaluate the effectiveness of these three fallback strategies as an ablation study.

\section{Experimental Setup}

\subsection{Datasets}
We evaluate all rerankers across 14 diverse datasets grouped into three families. Figure~\ref{fig:passage_lengths} summarizes the dataset statistics. We evaluate only the judged queries for each dataset. \textbf{TREC DL} covers web passage retrieval using MS MARCO queries from the 2019 \cite{craswell2020overview} and 2020 \cite{craswell2021overview} tracks, plus DL-Hard \cite{mackie2021dlhard}. \textbf{BEIR} includes five heterogeneous datasets \cite{thakur2021beir}: TREC-COVID, SciFact, DBpedia-Entity, FiQA, and ArguAna. \textbf{BRIGHT} includes six reasoning-intensive datasets \cite{su2025bright} (AoPS, Biology, Economics, Earth Science, Psychology, StackOverflow) that require complex multi-hop reasoning and feature exceptionally long passage tails, making them a rigorous test for context limits.

\subsection{Baselines}
We compare Jev against a comprehensive suite of established neural rerankers and frontier generative LLMs.

\textbf{Pointwise Rerankers.} These models score each document independently. We include representative models from three distinct architectural families:
\begin{itemize}
    \item \textbf{MiniLM-L6-v2:} A supervised cross-encoder run via Sentence Transformers with a maximum length of 512 tokens. It represents classical, highly efficient BERT-based pointwise reranking.
    \item \textbf{monoT5-3B-10K:} A supervised pointwise sequence-to-sequence model \cite{nogueira2020document} run via RankLLM \cite{sharifymoghaddam2025rankllm}, configured with a context limit of 512 tokens. It represents the standard encoder-decoder approach to relevance generation.
    \item \textbf{Qwen3-Reranker-0.6B:} A supervised pointwise reranker based on a causal language model \cite{yang2025qwen3}, supporting up to 8192 tokens. It represents modern, long-context causal LLMs fine-tuned specifically for pointwise scoring.
\end{itemize}

\textbf{Listwise and Setwise Rerankers.} These models evaluate multiple documents simultaneously to capture relative relevance.
\begin{itemize}
    \item \textbf{RankZephyr-7B:} A distilled listwise reranker \cite{pradeep2023rankzephyr} run via RankLLM and vLLM, using a window size of 20, a stride of 10, and 300 words per passage. It represents the state-of-the-art in open-weights listwise permutation generation.
    \item \textbf{Setwise (FLAN-T5-large):} A zero-shot setwise reranker \cite{zhuang2024setwise} that uses Heapsort over generation probabilities, truncating queries to 128 tokens and passages to fit within a 512-token context window. It represents the setwise selection paradigm using open-weights models.
\end{itemize}

\textbf{Multi-Paradigm Generative LLMs.} Frontier models can adapt to any paradigm.
\begin{itemize}
    \item \textbf{GPT-4o-mini:} A generative LLM baseline. Unlike the specialized models above, GPT-4o-mini can be applied to all categories. To isolate the effect of the model architecture from the prompt design, we evaluate GPT-4o-mini across all three paradigms by sending it the exact same \texttt{Noul} (pointwise binary), \texttt{Score} (pointwise graded), and \texttt{Choice} (list-based selection) instructions as Jev.
\end{itemize}

\subsection{Implementation Details}

\textbf{First-Stage Retrieval.} To ensure a strictly controlled comparison, every method reranks the exact same candidate list. We use Pyserini \cite{lin2021pyserini} to retrieve the top 100 candidates using BM25 on prebuilt indexes.

\noindent \textbf{Metrics and Statistical Protocol.} Our primary effectiveness metric is nDCG@10, computed using linear graded gain from the original qrels to maintain comparability with prior work. We also report MAP@10 for all the experiments of this paper in our repository. All metrics are calculated using \texttt{ir\_measures}. Ties are broken deterministically using the original BM25 rank. To determine statistical significance, we perform two-sided paired t-tests on per-query nDCG@10 scores, applying the Holm-Bonferroni correction ($p < 0.05$) within each dataset. We test each method against the BM25 baseline and test each Jev mode against the strongest baseline per dataset.

\noindent \textbf{Model Configurations.} For Jev, we use \texttt{jev-1.13.0}. For GPT-4o-mini, we use the \texttt{gpt-4o-mini-2024-07-18} snapshot with a temperature of 0 to ensure deterministic generation. All local supervised models are evaluated on a single GPU environment to measure comparable inference latency.

\noindent \textbf{Efficiency Measurements.} For efficiency, we measure the median reranking wall-clock latency per query in seconds. We isolate the reranking time by excluding BM25 retrieval, evaluation overhead, and model loading times. For hosted APIs (Jev and GPT-4o-mini), we report API calls per query, token usage, and the cost in USD per 1,000 queries based on list prices at the time of the run.

\section{Results}

\begin{table*}[t]
\centering
\caption{Effectiveness results (nDCG@10) comparing the three Jev modes against BM25, established neural rerankers, and GPT-4o-mini using the exact same prompts. \texttt{Choice} results reflect the Native mode (or Truncate where Native failed entirely). For each dataset, the best result is shown in \textbf{bold}, and the second-best is \underline{underlined}.}
\label{tab:main_results}
\renewcommand{\arraystretch}{1.25}
\setlength{\tabcolsep}{2pt}
\resizebox{\textwidth}{!}{
\begin{tabular}{clccc|cccccc|ccc}
\toprule
\textbf{Family} & \textbf{Dataset} & \textbf{Jev} & \textbf{Jev} & \textbf{Jev} & \textbf{BM25} & \textbf{MiniLM} & \textbf{monoT5} & \textbf{Qwen3} & \textbf{RankZephyr} & \textbf{Setwise} & \textbf{4o-mini} & \textbf{4o-mini} & \textbf{4o-mini} \\
 & & \textbf{Noul} & \textbf{Score} & \textbf{Choice} & & \textbf{L6} & \textbf{3B} & \textbf{0.6B} & \textbf{7B} & \textbf{FLAN-T5} & \textbf{Noul} & \textbf{Score} & \textbf{Choice} \\
\midrule

\multirow{3}{*}{\rotatebox[origin=c]{90}{TREC DL}}
& DL19 & 0.667 & 0.707 & 0.551 & 0.497 & \underline{0.732} & 0.722 & 0.701 & 0.731 & 0.669 & 0.731 & \textbf{0.748} & 0.518 \\
& DL20 & 0.672 & 0.686 & 0.548 & 0.488 & 0.679 & \underline{0.696} & 0.637 & \textbf{0.711} & 0.614 & 0.692 & 0.692 & 0.509 \\
& DL-Hard & 0.350 & 0.371 & 0.314 & 0.290 & \textbf{0.397} & \underline{0.393} & 0.354 & 0.383 & 0.366 & 0.354 & 0.364 & 0.296 \\

\midrule

\multirow{5}{*}{\rotatebox[origin=c]{90}{BEIR}}
& TREC-COVID & 0.792 & 0.835 & 0.652 & 0.595 & 0.745 & 0.794 & \textbf{0.860} & \underline{0.840} & 0.770 & 0.780 & 0.786 & 0.609 \\
& SciFact & \underline{0.788} & \textbf{0.795} & 0.775 & 0.679 & 0.687 & 0.765 & 0.767 & 0.759 & 0.626 & 0.731 & 0.740 & 0.712 \\
& DBpedia & 0.386 & 0.432 & 0.358 & 0.318 & \underline{0.442} & \textbf{0.448} & 0.428 & 0.441 & 0.414 & 0.380 & 0.404 & 0.336 \\
& FiQA & 0.406 & \underline{0.412} & 0.342 & 0.236 & 0.347 & \textbf{0.460} & 0.397 & 0.381 & 0.323 & 0.368 & 0.324 & 0.256 \\
& ArguAna & \underline{0.472} & 0.415 & 0.448 & 0.401 & 0.417 & 0.431 & \textbf{0.685} & 0.439 & 0.230 & 0.174 & 0.190 & 0.372 \\

\midrule

\multirow{6}{*}{\rotatebox[origin=c]{90}{BRIGHT}}
& AoPS & 0.049 & \underline{0.105} & 0.083 & 0.065 & 0.081 & 0.086 & \textbf{0.109} & 0.079 & 0.035 & 0.059 & 0.076 & 0.063 \\
& Biology & 0.274 & 0.285 & 0.227 & 0.181 & 0.118 & 0.102 & 0.169 & 0.242 & 0.145 & \underline{0.359} & \textbf{0.390} & 0.206 \\
& Economics & 0.207 & 0.226 & 0.194 & 0.165 & 0.123 & 0.128 & 0.153 & 0.179 & 0.120 & \textbf{0.251} & \underline{0.250} & 0.162 \\
& Earth Sci. & 0.315 & 0.339 & 0.287 & 0.281 & 0.186 & 0.232 & 0.276 & 0.298 & 0.161 & \underline{0.481} & \textbf{0.488} & 0.289 \\
& Psychology & 0.235 & 0.245 & 0.187 & 0.134 & 0.118 & 0.136 & 0.209 & 0.220 & 0.159 & \textbf{0.322} & \underline{0.314} & 0.138 \\
& StackOverflow & 0.185 & 0.193 & 0.159 & 0.162 & 0.122 & 0.112 & 0.189 & 0.155 & 0.087 & \textbf{0.255} & \underline{0.244} & 0.171 \\

\bottomrule
\end{tabular}
}
\end{table*}

In this section, we present the findings of our empirical evaluation. We address RQ1 and RQ2 in Section 5.1 by analyzing the effectiveness of different relevance evaluation paradigms across non-generative decision models (Jev), supervised neural baselines, and frontier generative LLMs (GPT-4o-mini). In Section 5.2, we delve deeper into the context limit failures observed during listwise evaluation. Finally, we answer RQ3 in Section 5.3 by comparing the latency and financial cost of these approaches.

\begin{figure}[t]
\centering
\includegraphics[width=1\textwidth]{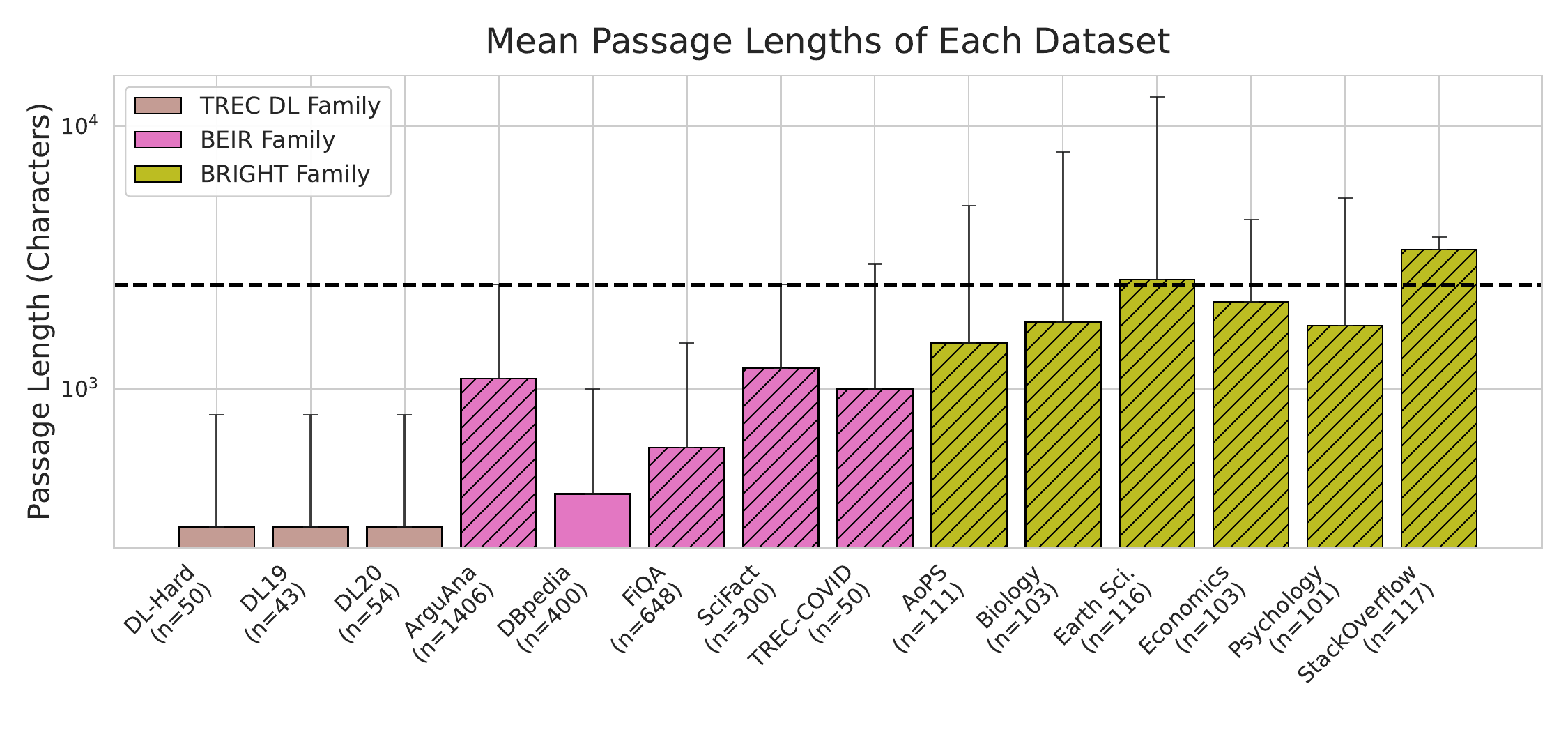}
\caption{Mean passage lengths across 14 datasets, with 95th-percentile error bars. The dashed line marks the safe per-passage budget ($\sim$2,500 chars) for a 100-passage context window. Hatched bars indicate datasets where native \texttt{Choice} failed due to context limits.}
\label{fig:passage_lengths}
\vspace{-1em}
\end{figure}

\subsection{Effectiveness of Reranking Paradigms}
Table~\ref{tab:main_results} presents the primary nDCG@10 effectiveness results across all 14 datasets. The table compares the three Jev modes (\texttt{Noul}, \texttt{Score}, \texttt{Choice}) against the BM25 baseline, five supervised neural models, and the three corresponding modes for GPT-4o-mini. A broad overview of the table reveals a clear hierarchy among the relevance evaluation paradigms: pointwise approaches (\texttt{Noul} and \texttt{Score}) consistently improve upon the BM25 baseline across almost all datasets, whereas list-based approaches (\texttt{Choice} and Setwise) frequently struggle or fail entirely. Specifically, Jev-\texttt{Score} significantly outperforms the BM25 baseline on 10 of the 14 datasets. Furthermore, on 6 of the 14 datasets, the performance difference between Jev-\texttt{Score} and the absolute strongest baseline for that dataset (either GPT-4o-mini or the best neural model) is not statistically significant, confirming that Jev can achieve top-tier effectiveness without the latency of autoregressive generation. Below, we break down these results to answer RQ1 and RQ2.

\begin{figure}[t]
\centering
\includegraphics[width=\textwidth]{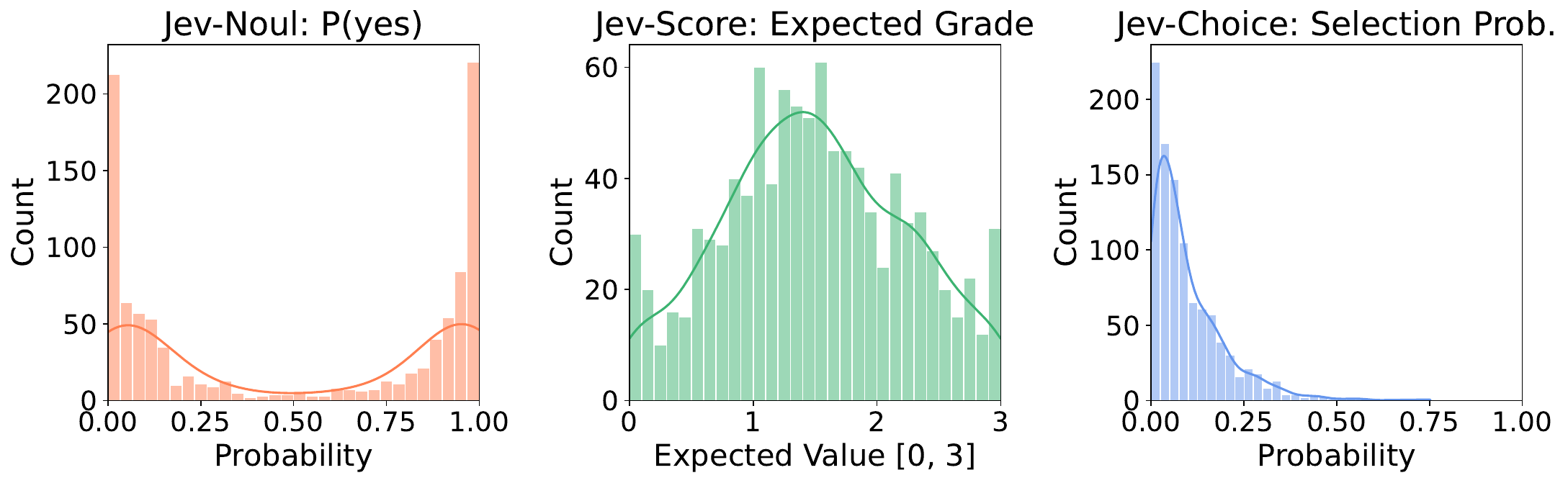}
\caption{Score distributions for the three Jev modes. \texttt{Noul} probabilities saturate near 0 and 1; \texttt{Score} returns expected values between levels (0 to 3); \texttt{Choice} exhibits a long tail and first-option bias.}
\label{fig:score_dist}
\vspace{-1em}
\end{figure}

\noindent \textbf{Finding 1. Pointwise graded scoring (\texttt{Score}) is the most effective paradigm for non-generative decision models, matching or exceeding the performance of frontier generative LLMs on many datasets.} As illustrated in Figure~\ref{fig:score_dist}, the \texttt{Score} mode returns an expected value over a graded scale, providing a much richer ranking signal than the binary probabilities of \texttt{Noul} or the skewed selection probabilities of \texttt{Choice}. Across almost all datasets in Table~\ref{tab:main_results}, Jev-\texttt{Score} outperforms the binary Jev-\texttt{Noul} and significantly outperforms Jev-\texttt{Choice}. Furthermore, Jev-\texttt{Score} achieves effectiveness comparable to, or even exceeding, the much more expensive GPT-4o-mini \texttt{Score} baseline. For instance, on TREC DL-Hard, Jev-\texttt{Score} achieves an nDCG@10 of 0.371, edging out GPT-4o-mini \texttt{Score} (0.364). On BEIR datasets like TREC-COVID, SciFact, and FiQA, Jev-\texttt{Score} consistently outperforms GPT-4o-mini \texttt{Score}. Conversely, GPT-4o-mini retains a strong advantage on reasoning-intensive BRIGHT datasets (e.g., Earth Science and Biology), suggesting that while non-generative grading is highly effective for standard reranking, autoregressive reasoning benefits complex multi-hop tasks.


\begin{table}[t]
\centering
\caption{Effectiveness (nDCG@10) of \texttt{Choice} fallback strategies. For datasets where Native \texttt{Choice} exceeded context limits, we compare Truncation and Tournament decomposition. The default Tournament mode failed on datasets with extreme long-tail passage lengths.}
\label{tab:choice_strategies}
\setlength{\tabcolsep}{3pt}
\begin{tabular}{lccc}
\toprule
\textbf{Dataset} & \textbf{Jev-Choice-Native} & \textbf{Jev-Choice-Truncate} & \textbf{Jev-Choice-Tournament} \\
\midrule
TREC DL19 & 0.551 & 0.543 & 0.623 \\
TREC DL20 & 0.548 & 0.541 & 0.615 \\
TREC DL-Hard & 0.314 & 0.317 & 0.355 \\
DBpedia & 0.358 & 0.354 & 0.408 \\
\midrule
TREC-COVID & \textit{Failed} & 0.652 & 0.750 \\
SciFact & \textit{Failed} & 0.775 & 0.800 \\
FiQA & \textit{Failed} & 0.342 & 0.397 \\
ArguAna & \textit{Failed} & 0.448 & 0.436 \\
AoPS & \textit{Failed} & 0.083 & 0.121 \\
Biology & \textit{Failed} & 0.227 & 0.240 \\
Economics & \textit{Failed} & 0.194 & \textit{Failed} \\
Earth Sci. & \textit{Failed} & 0.287 & \textit{Failed} \\
Psychology & \textit{Failed} & 0.187 & \textit{Failed} \\
StackOverflow & \textit{Failed} & 0.159 & \textit{Failed} \\
\bottomrule
\end{tabular}
\vspace{-1em}
\end{table}

\noindent \textbf{Finding 2. List-based selection (\texttt{Choice}) fails catastrophically across both non-generative and generative models, frequently degrading performance near or below the BM25 baseline.} The Setwise/\texttt{Choice} paradigm struggles severely. Jev-\texttt{Choice} is typically the worst-performing mode (e.g., on BRIGHT StackOverflow, Jev-\texttt{Choice} drops to 0.159 compared to BM25's 0.162, while Jev-\texttt{Score} reaches 0.193). Crucially, this failure is not unique to Jev. The GPT-4o-mini \texttt{Choice} baseline exhibits similarly poor performance across the board, scoring lower than BM25 on ArguAna, AoPS, and Economics, and barely outperforming it on Earth Science and Psychology despite aggressive truncation. The local Setwise (FLAN-T5) model also struggles heavily on BRIGHT. This indicates a fundamental failure of the listwise prompt paradigm itself over long contexts, likely driven by the lost-in-the-middle phenomenon, distraction by irrelevant documents, and the inherent difficulty of adhering to strict output formats when reasoning over 100 passages simultaneously.

\noindent \textbf{Finding 3. While supervised neural models retain an advantage on their in-domain datasets, Jev-\texttt{Score} matches or exceeds their performance on out-of-domain tasks.} The supervised neural models (monoT5, RankZephyr, Qwen3) remain strong baselines on the TREC DL datasets, which is expected given they were explicitly trained or distilled on MS MARCO. However, on out-of-domain datasets like SciFact, Jev-\texttt{Score} (0.795) exceeds monoT5 (0.765) and RankZephyr (0.759). On the complex BRIGHT Earth Science dataset, Jev-\texttt{Score} (0.339) substantially outperforms monoT5 (0.232) and RankZephyr (0.298), proving its robust zero-shot generalization capabilities across diverse retrieval domains.

\subsection{Context Limits and List-based Failures}
Beyond poor effectiveness, the \texttt{Choice} paradigm suffers from severe context limit vulnerabilities. A single native call containing 100 passages routinely exceeds Jev's default 64k token context limit on datasets with long passage tails. 

\noindent \textbf{Finding 4. Tournament decomposition provides an effectiveness boost over truncation, but fails to mitigate context limit errors on datasets with long-tail passage distributions.} To investigate whether fallback strategies could mitigate the context limit failures of native \texttt{Choice}, we evaluated truncation and tournament decomposition across the datasets. As shown in Table~\ref{tab:choice_strategies}, tournament decomposition successfully completed on 10 of the 14 datasets, providing a modest effectiveness boost over both truncation and the native approach. Because these strategies do not universally succeed across all datasets, we present them here as a separate set of experiments.

However, on datasets with extreme passage lengths (e.g., Earth Science, Economics, Psychology, and StackOverflow), tournament decomposition still failed. We found that shrinking the group size $w$ increases the total number of groups required to cover the top 100, which in turn forces the final tournament round to gather finalists from every group. For example, on BRIGHT Earth Science, a configuration of $w=8, m=1$ results in 13 finalists. Even when we tried highly restrictive configurations (e.g., $w=12, m=1$, which yields only 9 finalists), the final call still exceeded the context limit. This demonstrates that whether a call fits within the context limit depends on the combined length of its query and selected passages, not simply the number of finalists. As illustrated in Figure~\ref{fig:passage_lengths}, Earth Science features a mean passage length of 2,617 characters, which already exceeds the safe per-passage budget of $\sim$2,500 characters. Furthermore, even for datasets where the mean falls below the budget (e.g., SciFact, FiQA, and ArguAna), their long-tail distributions still exceed the limit. Consequently, native list-based \texttt{Choice} routinely overflows the context window, rendering it unviable for heterogeneous corpora.

\begin{figure}[t]
\centering
\includegraphics[width=0.7\textwidth]{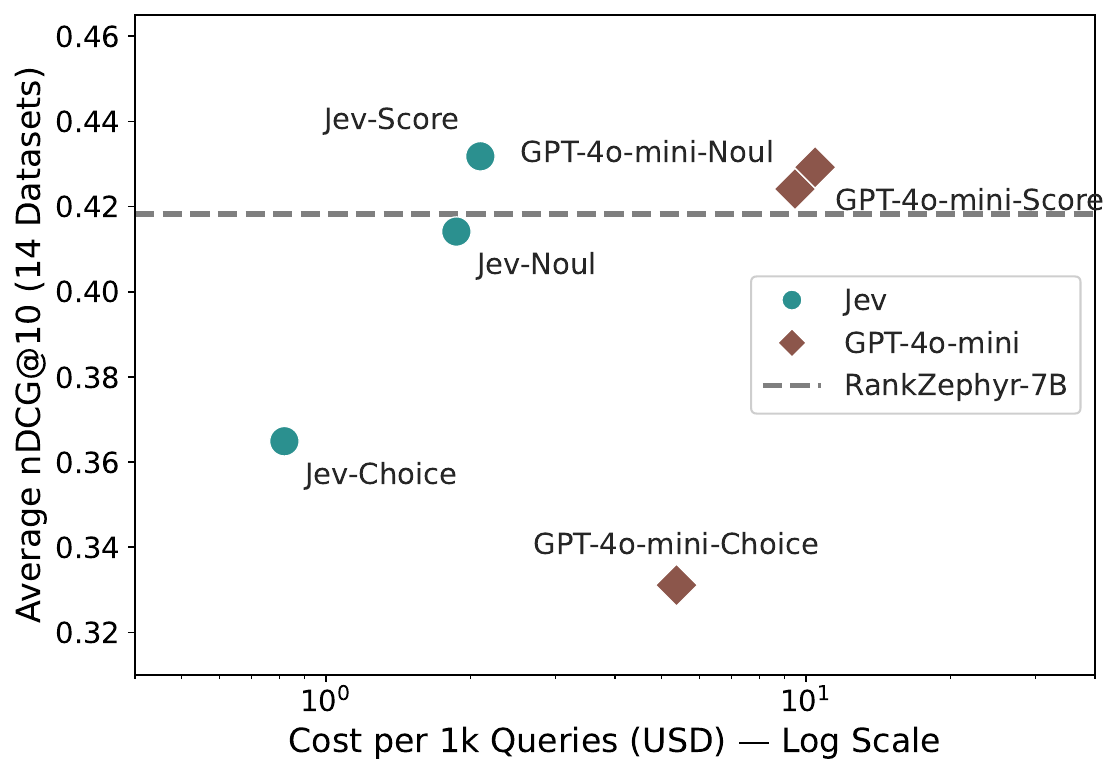}
\caption{Cost vs. Average nDCG@10 across 14 datasets. The dashed line marks the strongest local open-weights baseline (RankZephyr-7B). Jev-\texttt{Score} achieves comparable effectiveness to GPT-4o-mini at a fraction of the cost.}
\label{fig:pareto}
\vspace{-1em}
\end{figure}

\subsection{Efficiency Analysis}
Table~\ref{tab:efficiency} details the efficiency of the rerankers. Jev offers a massive latency and cost advantage over frontier generative LLMs. 

\noindent \textbf{Finding 5. Jev delivers up to a 28$\times$ speedup and up to an 80\% cost reduction compared to GPT-4o-mini while matching or exceeding its ranking effectiveness.} As visualized in the Pareto front in Figure~\ref{fig:pareto}, Jev fundamentally alters the trade-off between cost and effectiveness. While GPT-4o-mini requires between 7 and 50 seconds per query for the \texttt{Noul} and \texttt{Score} prompts and costs between \$3.94 and \$24.41 per 1,000 queries (scaling heavily with document length), Jev processes the exact same prompts in 0.6 to 0.8 seconds per query at a cost of roughly \$0.88 to \$4.78. This establishes decision models as a vastly more economical alternative to generative APIs.

\noindent \textbf{Finding 6. Jev's latency is highly competitive with heavy neural models, making it an efficient zero-shot option for production deployments.} While small cross-encoders like MiniLM are exceptionally fast (0.13s), Jev outpaces heavier neural models like RankZephyr-7B (21.65s) and Setwise FLAN-T5 (7.16s). By bypassing autoregressive decoding, Jev achieves the reasoning depth of large models without the latency penalty.

\begin{table}[t]
\centering
\caption{Efficiency metrics including median latency and cost per 1k queries. Neural model latencies are reported for a single GPU and do not incur API costs.}
\label{tab:efficiency}
\setlength{\tabcolsep}{12pt}
\begin{tabular}{lcc}
\toprule
\textbf{Method} & \textbf{Latency (s)} & \textbf{USD / 1k Queries} \\
\midrule
BM25 & $0.08 {\scriptstyle \pm 0.09}$ & N/A (Local) \\
MiniLM-L6-v2 & $0.13 {\scriptstyle \pm 0.07}$ & N/A (Local) \\
monoT5-3B-10K & $4.56 {\scriptstyle \pm 1.74}$ & N/A (Local) \\
Qwen3-Reranker-0.6B & $1.88 {\scriptstyle \pm 0.26}$ & N/A (Local) \\
RankZephyr-7B & $21.65 {\scriptstyle \pm 3.09}$ & N/A (Local) \\
Setwise (FLAN-T5) & $7.16 {\scriptstyle \pm 0.50}$ & N/A (Local) \\
\midrule
Jev-Noul & $0.72 {\scriptstyle \pm 0.05}$ & $\$1.87 {\scriptstyle \pm 1.12}$ \\
Jev-Score & $0.74 {\scriptstyle \pm 0.05}$ & $\$2.10 {\scriptstyle \pm 1.12}$ \\
Jev-Choice-Truncate & $0.23 {\scriptstyle \pm 0.08}$ & $\$0.82 {\scriptstyle \pm 0.25}$ \\
\midrule 
GPT-4o-mini Noul & $19.94 {\scriptstyle \pm 19.77}$ & $\$9.48 {\scriptstyle \pm 5.38}$ \\
GPT-4o-mini Score & $16.87 {\scriptstyle \pm 18.00}$ & $\$10.44 {\scriptstyle \pm 5.38}$ \\
GPT-4o-mini Choice & $1.61 {\scriptstyle \pm 0.83}$ & $\$5.37 {\scriptstyle \pm 3.54}$ \\
\bottomrule
\end{tabular}
\end{table}

\section{Conclusion}
\vspace{-0.5em}
In this paper, we presented the first large-scale benchmark of a non-generative ``System One'' decision model (Jev) for zero-shot document reranking across 14 datasets. By mapping established generative prompting paradigms to Jev's typed primitives, we investigated how non-generative architectures perform compared to supervised baselines and generative LLMs. Our findings indicate that Jev's pointwise graded mode (\texttt{Score}) matches or exceeds the effectiveness of frontier generative LLMs like GPT-4o-mini while offering significant reductions in latency and cost. Conversely, we exposed the inherent fragility of list-based (\texttt{Choice}) paradigms, which frequently fail on long-tail passage distributions across both generative and non-generative models due to context window overflows. Ultimately, this study provides empirical evidence that non-generative decision models offer a viable and highly efficient alternative to autoregressive LLMs for large-scale production reranking. To support future research and ensure the reproducibility of our findings, we release all artifacts associated with this study.

\subsection*{Limitations}
Several factors limit the generalizability of our findings. First, Jev is a closed, versioned API with undisclosed architecture and training data. We cannot definitively rule out test contamination on public datasets like MS MARCO, BEIR, or BRIGHT, though this threat applies equally to the GPT-4o-mini baseline. Second, we evaluate reranking strictly over a BM25 top-100 candidate list; results may differ when reranking deeper pools or candidates sourced from dense retrievers. Third, latency comparisons between GPUs and hosted APIs are inherently noisy, as API latency is subject to network overhead, rate pacing, and parallel worker configurations. Fourth, human relevance judgments are subject to assessor disagreement \cite{parry2025variations}, and the choice of evaluation metric can influence system rankings \cite{arabzadeh2025benchmarking}. Finally, the generative prompts we reproduced were originally designed for autoregressive models. While mapping them to typed primitives is a logical design choice, it remains an adaptation rather than a perfect 1:1 translation.

%
%
%

\clearpage

\bibliographystyle{splncs04}
\bibliography{references}
\end{document}